%% file: Dstoetapipipi.tex
\documentclass[prd,twocolumn,showpacs,amsmath,amssymb]{revtex4}
\usepackage{graphicx}
\usepackage{float}
\usepackage{epsfig,graphics,subfigure,psfrag,amsmath,amssymb}
\usepackage{lineno}
\usepackage{dcolumn}
\usepackage{bm}
\usepackage{overpic}
\usepackage{xspace}
\usepackage{rotating}
\usepackage{epstopdf}
\usepackage{makecell}
\usepackage{multirow}
\usepackage{dcolumn}
\usepackage{xcolor}
\usepackage{tikz}
\usepackage{array}  
\usepackage{booktabs}
\usepackage[colorlinks,linkcolor=blue, urlcolor=blue]{hyperref}
\usepackage{url}

\begin{document}

\title{\bf \boldmath Study of the Decay $D^{+}_{s}\rightarrow \pi^{+}\pi^{+}\pi^{-}\eta$ and Observation of the W-annihilation Decay $D^{+}_{s}\rightarrow a_0(980)^+\rho^0$}

\input{author_list}

\begin{abstract}
The decay $D^{+}_{s}\rightarrow \pi^{+}\pi^{+}\pi^{-}\eta$ is
observed for the first time, using $e^+e^-$ collision data
corresponding to an integrated luminosity of 6.32~fb$^{-1}$,
collected by the BESIII detector at center-of-mass energies
between 4.178 and 4.226~GeV.  The absolute branching fraction
for this decay is measured to be $\mathcal{B}(D^+_s \to \pi^+
\pi^+ \pi^- \eta) = (3.12\pm0.13_{\rm stat.}\pm0.09_{\rm syst.})$\%.  
The first amplitude analysis of this decay
reveals the sub-structures in $D^{+}_{s}\rightarrow
\pi^{+}\pi^{+}\pi^{-}\eta$ and determines the relative
fractions and the phases among these sub-structures.  The
dominant intermediate process is $D^{+}_{s}\rightarrow
a_1(1260)^+ \eta, a_1(1260)^+ \rightarrow \rho(770)^0\pi^+$
with a branching fraction of $(1.73 \pm 0.14_{\rm stat.} \pm
0.08_{\rm syst.})$\%.  We also observe the W-annihilation
process $D^{+}_{s}\rightarrow a_0(980)^+\rho(770)^0$,
$a_0(980)^+ \to \pi^+ \eta$ with a branching fraction of
$(0.21\pm0.08_{\rm stat.}\pm0.05_{\rm syst.})$\%, which is
larger than the branching fractions of other measured pure
W-annihilation decays by one order of magnitude.

\end{abstract}

\maketitle

Since the discovery of the charm quark in 1974, the hadronic decays of
charmed mesons have been extensively studied both experimentally and
theoretically.  However, making precise Standard Model predictions for
exclusive weak charm meson decays is rather difficult, because the
charm quark mass is not heavy enough to allow for a sensible heavy
quark expansion, as corrections of higher orders in $1/m_c$ become
very important, nor is it light enough for the application of chiral
perturbation theory~\cite{HYCHEN}.  Studies of hadronic $D^+_s$ decays
provide insight into its decay mechanisms. The Cabibbo-favored (CF)
hadronic $D^+_s$ decays mediated via a $c \to sW, W\to u \bar d$
transition, producing states with hidden strangeness ($s\bar s$ quarks
in $\eta$), have relatively large branching fractions (BFs), although
some of them have still not been measured.  The Particle Data Group
(PDG)~\cite{PDG} reports that the missing hadronic decays of $D^+_s$
with $\eta$ in the final state contribute a fraction of
($7.1\pm3.2$)\%.  Among them, the $D^+_s \to \pi^+\pi^+\pi^-\eta$
decay is expected to have a large BF, but until now it has not been
observed.

\begin{figure}[htbp]
	\centering
	\includegraphics[width=7.9cm]{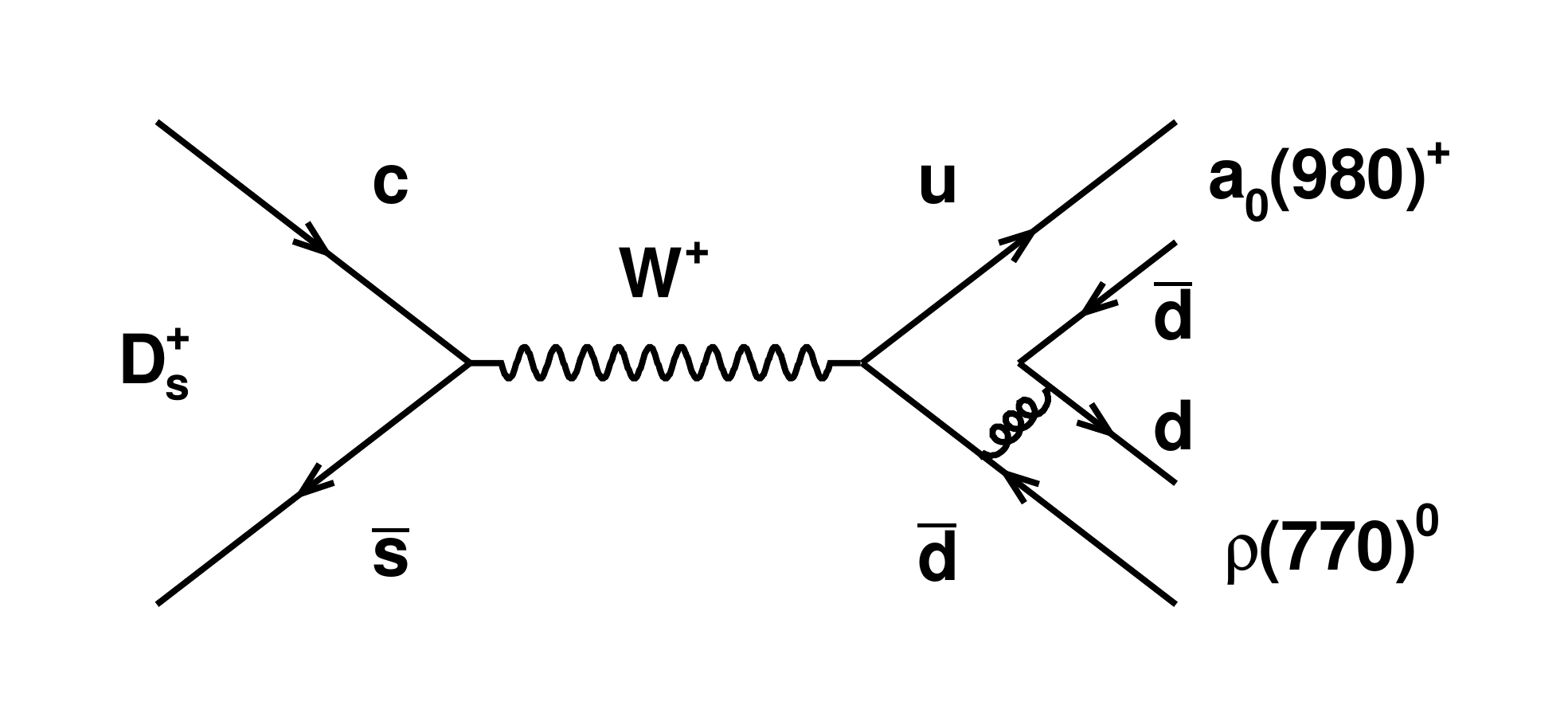}
	\caption{$D^+_s \to \pi^+\pi^+\pi^-\eta$ WA-topology diagrams, where the gluon lines can be connected with the quark lines in all possible configurations.}
		        \label{tp}
\end{figure}

The topological diagram analysis of the hadronic $D^+_s$
decays~\cite{tpd} reveals the importance of weak annihilation
contributions~\cite{tpfsi}. The amplitude of the long-distance weak
annihilation induced by final-state interactions may be comparable to
the tree amplitude~\cite{HYCHEN,tpfsi}.  The BF of the pure
W-annihilation (WA) process with a $PP$ final state, $D^+_s \to \pi^0
\pi^+$, is less than 0.037\%~\cite{pipi0}, and that of the $VP$ final
state, $D^+_s \to \rho(770)^0 \pi^+$, is 0.019\%~\cite{PDG}, while the
BF of the WA process with an $SP$ final state, $D^+_s \to
a_0(980)^{+(0)} \pi^{0(+)}, ~a_0(980)^{+(0)} \to \pi^{+(0)} \eta$, is
about 1.46\%~\cite{pipi0eta}. Here, $V$, $P$, and $S$ denote vector,
pseudoscalar, and scalar mesons, respectively.  Since the direct
production of the $a_0(980)\pi$ system via the $c\bar s \to W^+ \to
u\bar d \to a_0(980)\pi$ transition violates G-parity
conservation~\cite{gpar}, the WA process $D^+_s \to a_0^{+(0)}
\pi^{0(+)}$ is suppressed.  The origin of the abnormally large BF for
the $SP$ decay mode is still controversial.  Reference~\cite{Raquel}
argues that the $SP$ decay mode can be produced via internal emission
due to which $a_0(980)$ can be dynamically generated from $K\bar K$
final-state interactions in coupled channels. Reference~\cite{YuKuo}
claims that $D^+_s \to a_0(980)^{+(0)}\pi^{0(+)}$ receives the main
contribution from $D^+_s \to \rho(770)^{+(0)} \eta$ through a triangle
rescattering process. To date, there have been no theoretical or
experimental studies of the WA process with a $VS$ final state.  The
process $D^+_s \to a_0(980)^+\rho(770)^0$ can proceed via a WA
transition with G-parity conservation. The decay diagram for this decay is shown in Fig~\ref{tp}. A measurement of this process
is critical to distinguish various weak annihilation mechanisms.  In
addition, the underlying structure of the resonance $a_0(980)^+$ plays
an important role in the decay mechanism of the process $D^+_s \to
a_0(980)^+\rho(770)^0$.  However, the interpretation of $a_0(980)^+$
is still controversial~\cite{PDG}.  Studying the decay $D^+_s \to
a_0(980)^+\rho(770)^0$ can play a key role in understanding the nature
of $a_0(980)^+$.

A further motivation to study this decay arises from the tension
between theoretical calculations and experimental measurements of the
ratio of rates of semi-leptonic $B$ decays involving leptons from two
families.  The ratio of the BFs $R(D^*) = \mathcal{B}(B\to D^* \tau
\nu)/\mathcal{B}(B\to D^* l \nu)$ ($l = e, \mu$) averaged by HFLAV
is $0.295\pm0.011\pm0.008$. It differs from the Standard Model
prediction of $0.258\pm0.005$ by 2.6 standard
deviations~\cite{HFAG}. This suggests violation of the lepton flavor
universality.  However, experimental measurements at LHCb have a large
systematic uncertainty from the limited knowledge of the inclusive
$D^+_s \to \pi^+\pi^+\pi^- X$ decay~\cite{LFU}. A precise measurement
of the decay $D^+_s \to \pi^+\pi^+\pi^-\eta$ may provide useful input
to this problem.

This Letter reports the first observation of the $D^+_s \to \pi^+
\pi^+ \pi^- \eta$ decay, along with an amplitude analysis and a BF
measurement of this decay.  This analysis uses $e^+e^-$ collision data
samples corresponding to an integrated luminosity of 6.32~fb$^{-1}$
collected at the center-of-mass energies ($E_{\rm cm}$) between 4.178
and 4.226~GeV.  Throughout this Letter, charged-conjugate modes and
exchange symmetry of two identical $\pi^+$ are always implied.

The BESIII detector and the upgraded multi-gap resistive plate
chambers used in the time-of-flight systems are described in
Refs.~\cite{detector, chamber}, respectively.  Simulated data samples
produced with a {\sc geant4}-based~\cite{geant4} Monte Carlo (MC)
program, which includes the geometric description of the BESIII
detector and the detector response, are used to determine detection
efficiencies and to estimate backgrounds. The simulation models the
beam energy spread and initial state radiation (ISR) in the $e^+e^-$
annihilations with the generator {\sc kkmc}~\cite{KKMC}. The inclusive
MC sample includes the production of open charm processes modeled with
{\sc conexc}~\cite{conexc}, the ISR production of vector
charmonium(-like) states, and the continuum processes incorporated in
{\sc kkmc}~\cite{KKMC}. The known decay modes are modeled with {\sc
evtgen}~\cite{evtgen} using BFs taken from the
PDG~\cite{PDG}, and the remaining unknown charmonium decays are
modeled with {\sc lundcharm}~\cite{LundChar}. Final state
radiation~(FSR) from charged final-state particles is incorporated
using {\sc photos}~\cite{PHOTOS}.

We employ the double-tag (DT) technique~\cite{DT} to select $D^+_s
\to \pi^+\pi^+\pi^-\eta$ decays in $e^+e^- \to D^{*\pm}_sD^{\mp}_s,
D^{*\pm}_s \to \gamma D^{\pm}_s$ events.  The single-tag (ST) $D^-_s$
candidates are reconstructed from eight hadronic decay modes (tag
side): $D^-_s \to K^0_SK^-$, $K^+K^-\pi^-$, $K^0_SK^-\pi^0$,
$K^-K^+\pi^-\pi^0$, $K^0_SK^-\pi^+\pi^-$, $K^0_SK^+\pi^-\pi^-$,
$\pi^-\eta'$, and $K^-\pi^+\pi^-$.  These tag modes are combined to
perform an amplitude analysis.  A DT candidate is selected by
reconstructing the $D^+_s \to \pi^+\pi^+\pi^-\eta$ decay (signal side)
from the remaining particles that are not used in the ST
reconstruction.  Here, the $K^0_S$, $\pi^0$, $\eta$ and $\eta'$ mesons
are reconstructed from $K^0_S\to\pi^+\pi^-$, $\pi^0\to\gamma\gamma$,
$\eta\to\gamma\gamma$ and $\eta'\to\pi^+\pi^-\eta$ decays,
respectively.  The selection criteria for charged and neutral particle
candidates are identical to those used in Ref.~\cite{Dstokkpi}.  For
the decay mode $D^-_s \to K^-\pi^+\pi^-$, we exclude the di-pion mass
range [0.487, 0.511]~GeV/$c^2$ to avoid overlap with the $D^-_s \to
K^0_SK^-$ mode.

The invariant mass of the tag (signal) $D^{-(+)}_s$ candidate $M_{\rm
tag}$ ($M_{\rm sig}$) is required to be within the range [1.87,
2.06]~GeV/$c^2$.  We calculate the recoiling mass $M_{\rm rec}
= \{[E_{\rm cm}-(\vec{p}^2_{D^-_{s}} +
m^2_{D^-_s})^{1/2}]^2-|\vec{p}_{D^-_s}|^2\}^{1/2}$ in the $e^+e^-$
center-of-mass system, where $\vec{p}_{D^-_s}$ is the momentum of the
reconstructed $D^-_s$ and $m_{D^-_s}$ is the known mass of the $D^-_s$
meson~\cite{PDG}.  The value of $M_{\rm rec}$ is required to be in the
range [2.05, 2.18]~GeV/$c^2$ for the data sample collected at
4.178~GeV to suppress the non-$D^{*\pm}_sD^{\mp}_s$ events.  The
$M_{\rm rec}$ ranges for the other data samples are the same as those
in Ref.~\cite{Dstokkpipi0}.

A seven-constraint kinematic fit is applied to the $e^+e^- \to
D^{*\pm}_s D^{\mp}_s \to \gamma D^{+}_sD^{-}_s$ candidates, where
$D^-_s$ decays to one of the tag modes and $D^+_s$ decays to the
signal mode. In addition to the constraints of four-momentum
conservation in the $e^+e^-$ center-of-mass system, the invariant
masses of $\eta$, tag $D^-_s$, and $D^{*+}_s$ candidates are
constrained to their individual PDG values~\cite{PDG}.  If there are
multiple candidates (in $\approx15\%$ of the selected events) in an event,
the one with the minimum $\chi^2$ of the seven-constraint kinematic
fit is accepted.

To suppress the background from $D^+_s \to K^0_S \pi^+\eta$ decays, a
secondary vertex fit~\cite{svtx} is performed on the $\pi^+\pi^-$
pair.  If its invariant mass is in the range [0.487, 0.511]~GeV/$c^2$
and the flight distance between the interaction point (IP)~\cite{svtx}
and the decay point is two times greater than its uncertainty, we reject 
these candidates.  To suppress the background from the decay $D^+_s
\to \pi^+\eta'$ with $\eta' \to \pi^+\pi^-\eta$, we reject candidates
with $M_{\pi^+\pi^-\eta}<$ 1~GeV/$c^2$.  
To suppress the background where individual photons 
from random $\pi^0$'s feed into the $\eta \to \gamma \gamma$ reconstruction,
we define two invariant masses
$M(\gamma_{\eta}\gamma_{\pi^0})$ and $M(\gamma_{\eta}\gamma_{\rm
other})$, where $\gamma_{\eta}$, $\gamma_{\pi^0}$, and $\gamma_{\rm
other}$ denote the photon of $\eta$ from the signal side, the photon
of $\pi^0$ from the tag side, and the other photons including the
transition photon from $D^{*+(-)}_s$, respectively.  We reject
events with $M(\gamma_{\eta}\gamma_{\pi^0})$ or
$M(\gamma_{\eta}\gamma_{\rm other})$ in the range [0.115,
0.150]~GeV/$c^2$.

We further reduce the background by using a gradient-boosted decision
tree (BDT) implemented in the TMVA software package~\cite{TMVA}.  The
BDT takes four discriminating variables as inputs: the recoiling mass
of $D^*_s$, the momentum of the lower-energy photon from $\eta$, the
invariant mass of the two photons used to reconstruct $\eta$, and the
energy of the transition photon from $D^*_s$.  We place a requirement
on the output of the BDT to ensure the samples have a purity greater
than 85\%: $(87.0\pm3.8)\%$, $(85.6\pm4.9)\%$, and $(89.2\pm9.2)\%$ at
$E_{\rm cm} = 4.178$, 4.189-4.219, and 4.226~GeV, respectively.

An unbinned maximum likelihood method is adopted in the amplitude
analysis of the $D^{+}_{s}\rightarrow\pi^{+}\pi^{+}\pi^{-}\eta$ decay. The
likelihood function is constructed with a probability density
function (PDF) in which the momenta of the four final-state particles
are used as inputs.  The total likelihood is the product of
the likelihoods for all data samples.
The total amplitude is modeled as a coherent sum over all intermediate
processes $M(p_j)=\sum{\rho_ne^{i\phi_n}A_n(p_j)}$, where
$\rho_ne^{i\phi_n}$ is the coefficient of the $n^{\rm th}$
amplitude with magnitude $\rho_n$ and phase $\phi_n$. The $n^{\rm th}$
amplitude $A_n(p_j)$ is given by $A_n =
P^1_nP^2_nS_nF^1_nF^2_nF^{3}_n$, where the indices 1, 2 and 3
correspond to the two subsequent intermediate resonances and the
$D_s^+$ meson, $F^i_n$ is the Blatt-Weisskopf barrier
factor~\cite{Blatt, spin} and $P^i_n$ is the propagator of the
intermediate resonance.  
The function $S_n$ describes angular momentum
conservation in the decay and is constructed using the covariant
tensor formalism~\cite{spin}.  
The relativistic Breit-Wigner
(RBW)~\cite{RBW} function is used to describe the propagator for the
resonances $\eta(1405)$, $f_1(1420)$ and $a_1(1260)$.  The resonance
$\rho(770)^+$ is parameterized by the Gounaris-Sakurai~\cite{GSrho}
lineshape, and the resonances $a_0(980)$ and $f_0(980)$ are parameterized
by a coupled Flatt\'{e} formula, and the parameters are fixed to the
values given in Ref.~\cite{Flatte1} and Ref.~\cite{Flatte2},
respectively.  We use the same parameterization to describe $f_0(500)$
as Ref.~\cite{f0500}.  
The masses and widths of the
intermediate resonances, except for $a_0(980)$, $f_0(980)$ and
$f_0(500)$, are taken from Ref.~\cite{PDG}.

The background PDF
$B(p_j)$ is constructed from inclusive MC samples by using
a multi-dimensional kernel density estimator~\cite{RooNDKeysPdf} with five independent Lorentz
invariant variables ($M_{\pi^+\pi^+}$, $M_{\pi^+\pi^-}$,
$M_{\pi^+\eta}$, $M_{\pi^-\eta}$ and $M_{\pi^+\pi^-\eta}$).  As a
consequence, the combined PDF can be written as
\begin{equation}\nonumber
		\epsilon R_4\Bigg[f_s\frac{|M(p_j)|^2}{\int \epsilon|M(p_j)|^2R_4dp_j} + (1-f_s)\frac{B_{\epsilon}(p_j)}{\int \epsilon B_{\epsilon}(p_j) R_4dp_j}\Bigg], 
\end{equation}
where $\epsilon$ is the acceptance function determined with
phase-space (PHSP) MC samples generated with a uniform distribution of
the $D^+_s \to \pi^+ \pi^+ \pi^- \eta$ decay over PHSP, $B_{\epsilon}$
is $B/\epsilon$, and $R_4dp_j$ is the element of four-body PHSP.  The
normalization integral in the denominator is determined by a MC
technique as described in Ref.~\cite{D0tokkpipi}.

\begin{figure}[htbp]
	\flushleft
	\includegraphics[width=8.65cm]{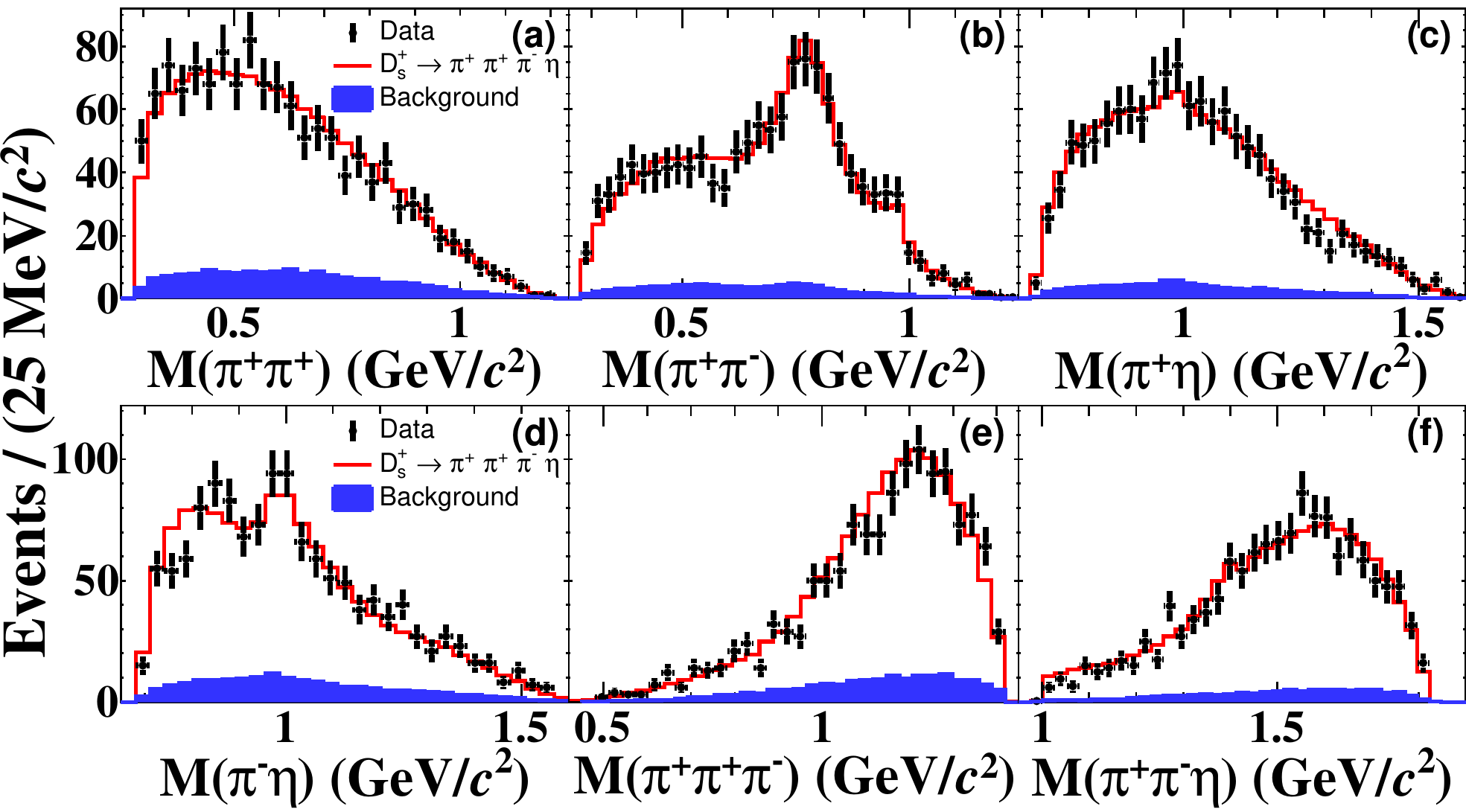}
	\caption{
          The (a) $M_{\pi^+\pi^+}$, (b) $M_{\pi^+\pi^-}$, (c) $M_{\pi^+\eta}$,
          (d) $M_{\pi^-\eta}$, (e) $M_{\pi^+\pi^+\pi^-}$ and (f) $M_{\pi^+\pi^-\eta}$ projections for
          data with the best fit superimposed. 
	  The dots with error bars, the red histograms and the blue histograms
	    are data from all samples, the total fit and the background
	      contribution estimated with inclusive MC samples, respectively.
	        Since the two $\pi^+$'s in the final state are identical particles, they have a symmetric wave function in some states. The projections of $M_{\pi^+\pi^-}$, $M_{\pi^+\pi^-\eta}$ and
		  $M_{\pi^+\eta}$ are added for the two $\pi^+$'s.}
		  	\label{pwa_plot}
\end{figure}

In the initial amplitude fit, we include a few obvious
components. Then, further amplitudes are added one at a time to the
fit, and the statistical significance of the new amplitude is
calculated with the change of the log-likelihood, after taking the
change of the degrees of freedom into account.  Only amplitudes with
significance larger than 5$\,\sigma$ are chosen for the optimal set.
The dominant CF amplitude to this final state is $D^+_s
\to a_1(1260)^+\eta,~ a_1(1260) \to [\rho(770)^0 \pi^+]_S$, where the
subscript $S$ 
means that the angular momentum of $\rho^0\pi^+$ combination is zero ($S$-wave).
Thus, we choose this amplitude as the reference and its
phase is fixed to 0.  The $D^+_s \to a_0(980)^+\rho(770)^0$,
$a_0(980)^+\to\pi^+\eta$ decay is observed with a significance larger
than $\,7\sigma$.  We also consider some possible amplitudes involving
the resonances $f_0(500)$, $f_0(980)$, $f_1(1285)$, $\eta(1295)$,
$\eta(1405)$, $\eta(1475)$, $f_1(1420)$, $f_1(1510)$ and $\pi(1300)$,
as well as non-resonant components.  Moreover, charge conjugation for
$D^+_s \to \eta(1405)(f_1(1420))\pi^+$ with $\eta(1405)(f_1(1420)) \to
a_0(980)^+ \pi^-$ and 
$\eta(1405)(f_1(1420)) \to a_0(980)^- \pi^+$ requires their magnitudes
and phases to be the same.  Finally, eleven amplitudes are retained in
the nominal fit, as listed in Table~\ref{tab:res}.  The mass
projections of the fit are shown in Fig.~\ref{pwa_plot}.  For the
$n^{\rm th}$ component, its contribution relative to the total BF is
quantified by the fit fraction (FF) defined by FF$_n = \int |\rho_n
A_n(p_j)|^2R_4\,dp_j/\int |M|^2 R_4\,dp_j$.  The final amplitudes, their
phases and FFs are listed in Table~\ref{tab:res}. The sum of the FFs
of all the components is ($95.0\pm4.9$)\%.  A $\chi^2$ value is
calculated to quantify the quality of the fit, as defined in
Ref.~\cite{D0tokkpipi}. The goodness of fit is $\chi^2$/NDOF =
153.2/133 = 1.15 and the p-value is 11.1\%.  In addition, 300 sets of signal MC samples with the
same size as the data samples are generated to validate the fit
performance, as described in Ref.~\cite{Dstokkpi}.  No significant
bias is found in our fit.

We determine the systematic uncertainties by taking the differences
between the values of $\phi_n$ and FF$_n$ found by the nominal fit and
those found from fit variations. The masses and widths of the
intermediate states are varied by $\pm 1\,\sigma$~\cite{PDG}. The
masses and coupling constants of $a_0(980)$ and $f_0(980)$ are varied
within the uncertainties reported in Ref.~\cite{Flatte1} and
Ref.~\cite{Flatte2}, respectively.  The barrier radii for $D^+_s$ and
the other intermediate states are varied by $\pm1$~GeV$^{-1}$.  The
uncertainties related to lineshape are estimated by using an
alternative RBW function for $f_0(500)$ with mass and width fixed to
526~MeV/$c^2$ and 535~MeV~\cite{f0500rbw}.  The uncertainties from
detector effects are investigated with the same method as described in
Ref.~\cite{Dstokkpi}.  The uncertainty related to the background is
estimated by varying the background yield within statistical
uncertainty, and constructing the background PDF with the other five
independent variables.  The total uncertainties are obtained by adding
all the contributions in quadrature, and are listed in
Table~\ref{tab:res}.

\begin{table}[htbp]
        \caption{Phases and FFs for various intermediate processes. The first and the second uncertainties are statistical and systematic, respectively.}
	        \centering
	        \setlength{\tabcolsep}{0.6mm}{
        \small
       \begin{tabular}{lcr}
       \hline
       Amplitude  &Phase  &\multicolumn{1}{c}{FF(\%)}  \\
      \hline
      \specialrule{0em}{0.5pt}{0.6pt}
    $a_1(1260)^+(\rho(770)^0\pi^+)\eta$ &0.0(fixed)             &$55.4\pm3.9\pm2.0$                  \\       
    $a_1(1260)^+(f_0(500)\pi^+)\eta$    &$5.0\pm0.1\pm0.1$      &$8.1\pm1.9\pm2.1$                   \\
    $a_0(980)^+\rho(770)^0$             &$2.5\pm0.1\pm0.1$      &$6.7\pm2.5\pm1.5$                   \\
    $\eta(1405)(a_0(980)^-\pi^+)\pi^+$  &$0.2\pm0.2\pm0.1$      &$0.7\pm0.2\pm0.1$                   \\
    $\eta(1405)(a_0(980)^+\pi^-)\pi^+$  &$0.2\pm0.2\pm0.1$      &$0.7\pm0.2\pm0.1$                   \\
    $f_1(1420)(a_0(980)^-\pi^+)\pi^+$   &$4.3\pm0.2\pm0.4$      &$1.9\pm0.5\pm0.3$                   \\
    $f_1(1420)(a_0(980)^+\pi^-)\pi^+$   &$4.3\pm0.2\pm0.4$      &$1.7\pm0.5\pm0.3$                   \\
    $[a_0(980)^-\pi^+]_S\pi^+$          &$0.1\pm0.2\pm0.2$      &$5.1\pm1.2\pm0.9$                   \\
    $[a_0(980)^+\pi^-]_S\pi^+$          &$0.1\pm0.2\pm0.2$      &$3.4\pm0.8\pm0.6$                   \\
    $[f_0(980)\eta]_S\pi^+$             &$1.4\pm0.2\pm0.3$      &$6.2\pm1.7\pm0.9$                   \\
    $[f_0(500)\eta]_S\pi^+$             &$2.5\pm0.2\pm0.3$      &$12.7\pm2.6\pm2.0$                  \\
																										         \hline
																									         \end{tabular}}
																								          \label{tab:res}
\end{table}

\begin{figure}[htbp]
	\flushleft
	\includegraphics[width=8.65cm]{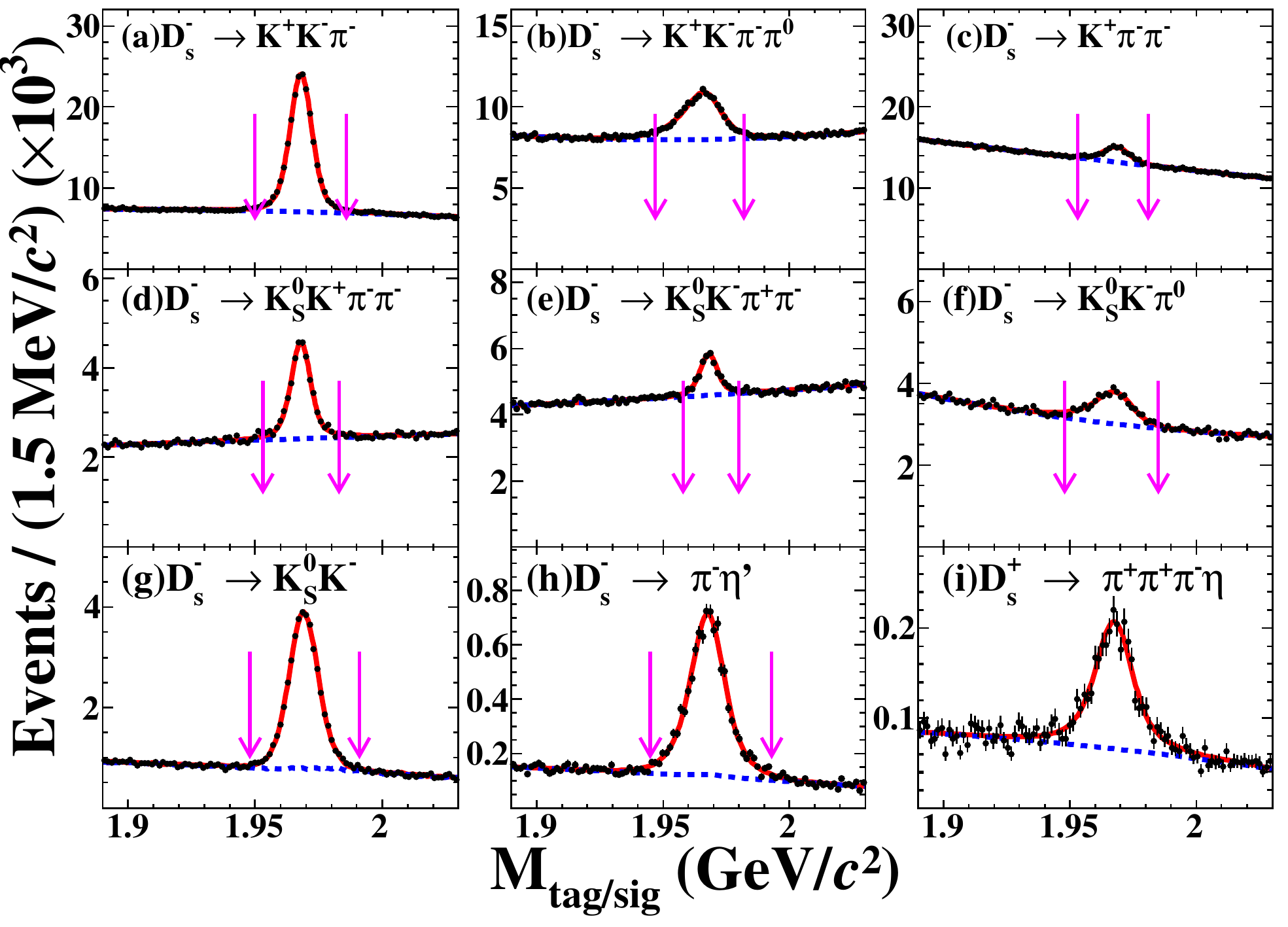}
	\caption{Fits to (a)-(h) the $M_{\rm tag}$ distributions of
  the ST candidates and (i) the $M_{\rm sig}$ distribution of
   the signal candidates. The dots with error bars are data
   from all samples. The red solid lines are the fits. The blue
   dashed lines are the fitted background shapes. The pair of
   pink arrows mark the chosen signal region. In the plots for 
   $D^-_s \to K^0_SK^-$ and $D^-_s \to \pi^-\eta'$ decays, the blue dashed lines
   include contributions from $D^- \to K^0_S\pi^-$ and $D^-_s\to \pi^+\pi^-\pi^-\eta$
	backgrounds, respectively.}
\label{fig:8tag}
\end{figure}

Further, we measure the BF of $D^+_s \to \pi^+ \pi^+ \pi^- \eta$ with
the DT technique.  To improve the statistical precision, the DT
candidates are selected without reconstructing the transition photon
from $D^{*}_s$ and without the BDT requirement.  We use the
same eight tag modes as used in the amplitude analysis.  For each tag
mode, if there are multiple tag $D^-_s$ candidates, the one with
$M_{\rm rec}$ closest to the known mass of $D^{*}_s$~\cite{PDG} is
retained. For each tag mode, a DT candidate with the average mass
$(M_{\rm sig}+M_{\rm tag})/2$ closest to the known mass of
$D_s$~\cite{PDG} is retained.  The ST yield ($Y_{\rm tag}$) and DT
yield ($Y_{\rm sig}$) in data are determined from fits to the $M_{\rm
tag}$ and $M_{\rm sig}$ distributions, respectively, as shown in
Fig.~\ref{fig:8tag}. The signal shape is modeled with the MC-simulated
shape convolved with a Gaussian function, and the background is
parameterized as a second-order Chebyshev function.

These fits result in a total ST yield of $Y_{\rm tag}=479, 093\pm1952$
and a signal yield of $Y_{\rm sig}=2139\pm78$ events. An updated inclusive MC
sample based on our amplitude analysis results is used to determine
the ST efficiencies ($\epsilon^i_{\rm ST}$) and DT efficiencies
($\epsilon^i_{\rm DT}$).  Inserting these numbers in
$\mathcal{B}(D^+_s \to \pi^+ \pi^+ \pi^- \eta) = Y_{\rm
sig}/(\mathcal{B}(\eta \to \gamma\gamma) \times
\Sigma_{i,\alpha}Y^{i,\alpha}_{\rm tag}\epsilon^{i,\alpha}_{\rm
DT}/\epsilon^{i,\alpha}_{\rm ST})$, where $i$ denotes the $i^{\rm
th}$ tag mode and $\alpha$ denotes the $\alpha^{\rm th}$ center-of-mass
energy point, we obtain $\mathcal{B}(D^+_s \to \pi^+ \pi^+ \pi^-
\eta)=(3.12\pm0.13)$\%, where the uncertainty is statistical only.

The systematic uncertainties for the BF measurement are described
next. The uncertainty of the signal yield and total ST yield is
assigned to be 1.4\% by examining the changes of the fit yields when
varying the signal and background shapes.  The $\pi^{\pm}$ tracking
(PID) efficiencies are studied using samples of $e^+e^- \to K^+ K^-
\pi^+ \pi^-$ ($e^+e^- \to K^+ K^- \pi^+ \pi^-$ and $\pi^+ \pi^- \pi^+
\pi^-$) events.  The corresponding systematic uncertainties are
estimated as 0.3\% (0.4\%).  The uncertainty due to the $\eta$
reconstruction efficiency is 2.0\%~\cite{pipi0eta}. The
uncertainty from the amplitude model is estimated to be 0.4\%, which
is the change of signal efficiency when the parameters are varied
according to the covariance matrix in the nominal amplitude fit.  The
uncertainty due to MC simulation sample size is 0.3\%, and that from
the BF of $\mathcal{B}(\eta \to \gamma\gamma)$ is 0.5\%~\cite{PDG}.
Adding these uncertainties in quadrature gives a total systematic
uncertainty of 2.9\%.

In summary, using $e^+e^-$ collision data equivalent to an
integrated luminosity of 6.32~fb$^{-1}$ recorded with the BESIII
detector at $E_{\rm cm}$= 4.178-4.226~GeV, we observe the $D^{+}_{s}
\to \pi^{+}\pi^{+}\pi^{-}\eta$ decay for the first time. The absolute
BF of this decay is measured to be $\mathcal {B}(D^{+}_{s} \to
\pi^{+}\pi^{+}\pi^{-}\eta)$ = ($3.12\pm0.13_{\rm stat.}\pm0.09_{\rm
syst.}$)\%.  The first amplitude analysis of this decay is also
performed. The obtained intermediate processes, phases and FFs are
summarized in Table~\ref{tab:res}.  The BFs for the intermediate
processes are calculated with $\mathcal {B}_n$ = FF$_n \times \mathcal
{B}(D^+_s\to \pi^+\pi^+\pi^-\eta)$.  The $D^{+}_{s} \to
a_1(1260)^+\eta,~ a_1(1260)^+ \to \rho(770)^0\pi^+$ decay is dominant
with a BF of ($1.73\pm0.14_{\rm stat.}\pm0.08_{\rm syst.}$)\%.  Our
results offer critical input for estimating the $D^+_s \to \pi^+ \pi^+
\pi^- X$ background contribution in tests of the lepton flavor
universality with semileptonic $B$ decays.
Assuming that $\mathcal{B}(a_1(1260)^+ \rightarrow \rho(770)^+\pi^0) = \mathcal{B}(a_1(1260)^+ \rightarrow \rho(770)^0\pi^+)$, the BF of $D^+_s\to \pi^+\pi^0\pi^0\eta$ decay is expected to be comparable to the one of $D^+_s\to \pi^+\pi^+\pi^-\eta$. In this case, the missing inclusive hadronic $\eta$ decay fraction of $D^+_s$ is reduced to $(2.3\pm3.2)$\%, 
thereby indicating that there is no large room for unobserved exclusive $D^+_s \to \eta X$ decays.
Furthermore, we observe the WA decay $D^{+}_{s} \to a_0(980)^+\rho(770)^0, ~a_0(980)^+ \to \pi^+ \eta$ with BF of ($0.21\pm0.08_{\rm stat.}\pm0.05_{\rm syst.})$\%.
This BF and the one of the decay $D^{+}_{s} \to a_0(980)^+\pi^0$ obtained in Ref.~\cite{pipi0eta} are both larger than those of the 
pure WA processes $D^{+}_{s} \to \rho(770)^0\pi^+$ and $D^{+}_{s} \to \pi^0\pi^+$ by one order of magnitude.
These measurements indicate that long-distance weak annihilation may play an essential role, and provide
a good opportunity to study the final-state rescattering in the WA process~\cite{HYCHEN, Raquel, YuKuo}.


\section*{ACKNOWLEDGMENTS}
The BESIII collaboration thanks the staff of BEPCII and the IHEP computing center for their strong support. This work is supported in part by National Key Research and Development Program of China under Contracts Nos. 2020YFA0406400, 2020YFA0406300; National Natural Science Foundation of China (NSFC) under Contracts Nos. 11625523, 11635010, 11735014, 11822506, 11835012, 11935015, 11935016, 11935018, 11961141012; the Chinese Academy of Sciences (CAS) Large-Scale Scientific Facility Program; Joint Large-Scale Scientific Facility Funds of the NSFC and CAS under Contracts Nos. U1732263, U1832107, U1832207; CAS Key Research Program of Frontier Sciences under Contract No. QYZDJ-SSW-SLH040; 100 Talents Program of CAS; INPAC and Shanghai Key Laboratory for Particle Physics and Cosmology; ERC under Contract No. 758462; European Union Horizon 2020 research and innovation programme under Contract No. Marie Sklodowska-Curie grant agreement No 894790; German Research Foundation DFG under Contracts Nos. 443159800, Collaborative Research Center CRC 1044, FOR 2359, GRK 214; Istituto Nazionale di Fisica Nucleare, Italy; Ministry of Development of Turkey under Contract No. DPT2006K-120470; National Science and Technology fund; Olle Engkvist Foundation under Contract No. 200-0605; STFC (United Kingdom); The Knut and Alice Wallenberg Foundation (Sweden) under Contract No. 2016.0157; The Royal Society, UK under Contracts Nos. DH140054, DH160214; The Swedish Research Council; U. S. Department of Energy under Contracts Nos. DE-FG02-05ER41374, DE-SC-0012069.

\end{document}

%% file: author_list.tex
\author{
\begin{small}
\begin{center}
	M.~Ablikim$^{1}$, M.~N.~Achasov$^{10,c}$, P.~Adlarson$^{67}$, S.~Ahmed$^{15}$, M.~Albrecht$^{4}$, R.~Aliberti$^{28}$, A.~Amoroso$^{66A,66C}$, M.~R.~An$^{32}$, Q.~An$^{63,49}$, X.~H.~Bai$^{57}$
% ...
	\\
	\vspace{0.2cm}
	(BESIII Collaboration)\\
	\vspace{0.2cm} {\it
	$^{1}$ Institute of High Energy Physics, Beijing 100049, People's Republic of China\\
% ...
$^{4}$ Bochum Ruhr-University, D-44780 Bochum, Germany\\
% ...
$^{10}$ G.I. Budker Institute of Nuclear Physics SB RAS (BINP), Novosibirsk 630090, Russia\\
% ...
$^{15}$ Helmholtz Institute Mainz, Johann-Joachim-Becher-Weg 45, D-55099 Mainz, Germany\\
% ...
$^{28}$ Johannes Gutenberg University of Mainz, Johann-Joachim-Becher-Weg 45, D-55099 Mainz, Germany\\
% ...
$^{32}$ Liaoning Normal University, Dalian 116029, People's Republic of China\\
% ...
$^{49}$ State Key Laboratory of Particle Detection and Electronics, Beijing 100049, Hefei 230026, People's Republic of China\\
% ...
$^{57}$ University of Jinan, Jinan 250022, People's Republic of China\\
% ...
$^{63}$ University of Science and Technology of China, Hefei 230026, People's Republic of China\\
% ...
$^{66}$ University of Turin and INFN, (A)University of Turin, I-10125, Turin, Italy; (B)University of Eastern Piedmont, I-15121, Alessandria, Italy; (C)INFN, I-10125, Turin, Italy\\
	$^{67}$ Uppsala University, Box 516, SE-75120 Uppsala, Sweden\\
% ...
$^{c}$ Also at the Novosibirsk State University, Novosibirsk, 630090, Russia\\
% ...
	}
\end{center}
\vspace{0.4cm}
\end{small}
}
\noaffiliation{}